\documentclass[11pt]{article}
\usepackage{amsfonts,url,epsfig,float}
\usepackage{geometry} 
\usepackage{sectsty} 
\usepackage[normalem]{ulem}
\usepackage{xcolor}

\sectionfont{\sffamily\bfseries\upshape\large}
\subsectionfont{\sffamily\bfseries\upshape\normalsize} 
\subsubsectionfont{\sffamily\mdseries\upshape\normalsize}
\makeatletter
\renewcommand\@seccntformat[1]{\csname the#1\endcsname.\quad}
\makeatother\renewcommand{\bibitem}{\vskip 2pt\par\hangindent\parindent\hskip-\parindent}

\makeatletter
\def\@maketitle{%
  \begin{center}%
  \let \footnote \thanks
    {\large \@title \par}%
    {\normalsize
      \begin{tabular}[t]{c}%
        \@author
      \end{tabular}\par}%
    {\small \@date}%
  \end{center}%
}
\makeatother

\newcommand{\btR}{\vspace{-.25in}\begin{quotation}\begin{small}\noindent\begin{verbatim}}

\title{\bf Causal quartets:  Different ways to attain the same average treatment effect\footnote{We thank Dan Goldstein, Stephen Stigler, and Howard Wainer for helpful comments and the U.S. Office of Naval Research and Institute of Education Sciences for partial support of this work.}\vspace{.1in}}

\author{Andrew Gelman, Jessica Hullman, and Lauren Kennedy\vspace{.1in}}

\date{22 Feb 2023\vspace{-.2in}}

\begin{document}\sloppy
\maketitle

\begin{abstract}
The average causal effect can often be best understood in the context of its variation.  We demonstrate with two sets of four graphs, all of which represent the same average effect but with much different patterns of heterogeneity.  As with the famous correlation quartet of Anscombe (1973), these graphs dramatize the way in which real-world variation can be more complex than simple numerical summaries.  The graphs also give insight into why the average effect is often much smaller than anticipated.
\end{abstract}

\section{Given that real-world effects vary, and statistics is the study of variation, why does the causal inference literature focus on average effects?}

Causal inference in statistics and economics focuses on the average causal effect.  The purpose of this paper is to raise awareness of different patterns of heterogeneous causal effects:  examples where the average effect does not tell the whole story.

Given that real-world effects vary, and statistics is the study of variation, it seems obvious to look at the variation of causal effects across different populations, different scenarios, different time frames, etc. We are not alone in advocating for the value of seeking to understand sources of variation: other authors, such as Baribalt et al.\ (2018), Bryan, Tipton, and Yeager (2021), and Yarkoni (2022), have argued for the importance of varying effects, both for theoretical understanding and practical decision-making. 
Indeed, the very phrase ``average causal effect'' implicitly considers how the effect might vary; otherwise one could simply say ``causal effect'' without the modifier. 

Perhaps surprisingly, though, much of the literature of statistics and econometrics focuses on the estimation of average causal effects without much discussion of variation.  Before proceeding to discuss the importance of varying treatment effects, it behooves us to consider why there has been such an interest in averages.

There are several good reasons for the traditional approach of considering the treatment effect to be a single parameter to be estimated:
\begin{itemize}
\item In a randomized experiment, the average difference comparing treatment and control groups yields an unbiased estimate of the sample average treatment effect.  It makes sense to study this average effect, as this is what can be estimated from the data.
\item More generally, under different assumptions in observational studies, various local average treatment effects are what can be identified (Imbens and Angrist, 1994).  
\item When estimating a causal model using linear regression without interactions, so the coefficient of the treatment variable represents the causal effect.  In the presence of varying treatment effects, this coefficient represents an average treatment effect, in the same way that fitting a linear model to nonlinear data can be considered to estimate some sort of average regression line. Hence it can make sense to speak of ``the'' causal effect in the same way that we would speak of ``the'' regression coefficient $\beta$, as representing a single parameter in a model or a population average quantity.
\item Interactions can be hard to estimate; indeed, under some reasonable assumptions you need 16 times the sample size to estimate an interaction than to estimate a main effect (Gelman, 2018a).  Thus it can make sense to fit a model with constant treatment effect even if you think there may be interactions in reality.
\item Under the assumption of a constant treatment effect (the ``Fisher null hypothesis''), it is possible to obtain exact confidence intervals for randomized experiments.
\end{itemize}
For all these reasons, along with the convenience of single-number summaries, it has become standard practice either to fit a model assuming a constant treatment effect or to aggregate to obtain an estimated average treatment effect when fitting models in which effects vary; see, for example, Hill (2011) and Wager and Athey (2018).

That all said, we have become convinced through work in many application areas that thinking about varying effects can be essential for understanding causal inference, and consequently for making decisions based on estimates, such as in implementing policies or interventions beyond the lab. In this article we present causal quartets as a graphical tool for helping reform how we think about effects. Sections \ref{quartets} and \ref{practical} demonstrate and explain the value of such tools. Section \ref{package} presents a software package that researchers or consumers of research can use to create causal quartets in order to reflect on their own research goals or interrogate effects in the literature. In Section \ref{discussion} we discuss implications of treatment-effect heterogeneity for statistical practice in the context of the reasons discussed above for traditionally focusing on the average.

\section{Two causal quartets}\label{quartets}

\subsection{Plots of latent causal effects}

We dramatize variation in causal effects with two ``quartets'':  sets of four plots with the same average effect but much different patterns of individual effects.  All the displays plot the causal effect vs.\ a hypothetical individual-level predictor, $x$.  The first quartet shows examples of unpredictable or random variation, so that $x$ is essentially just an index of units.  The second quartet shows effects that vary as different systematic functions of $x$.  More generally, $x$ could represent different types of units and could be an observed predictor or a latent quantity.  The quartets are conceptual plots of different scenarios, not direct graphs of data.

\begin{figure}
\centerline{\includegraphics[width=.8\textwidth]{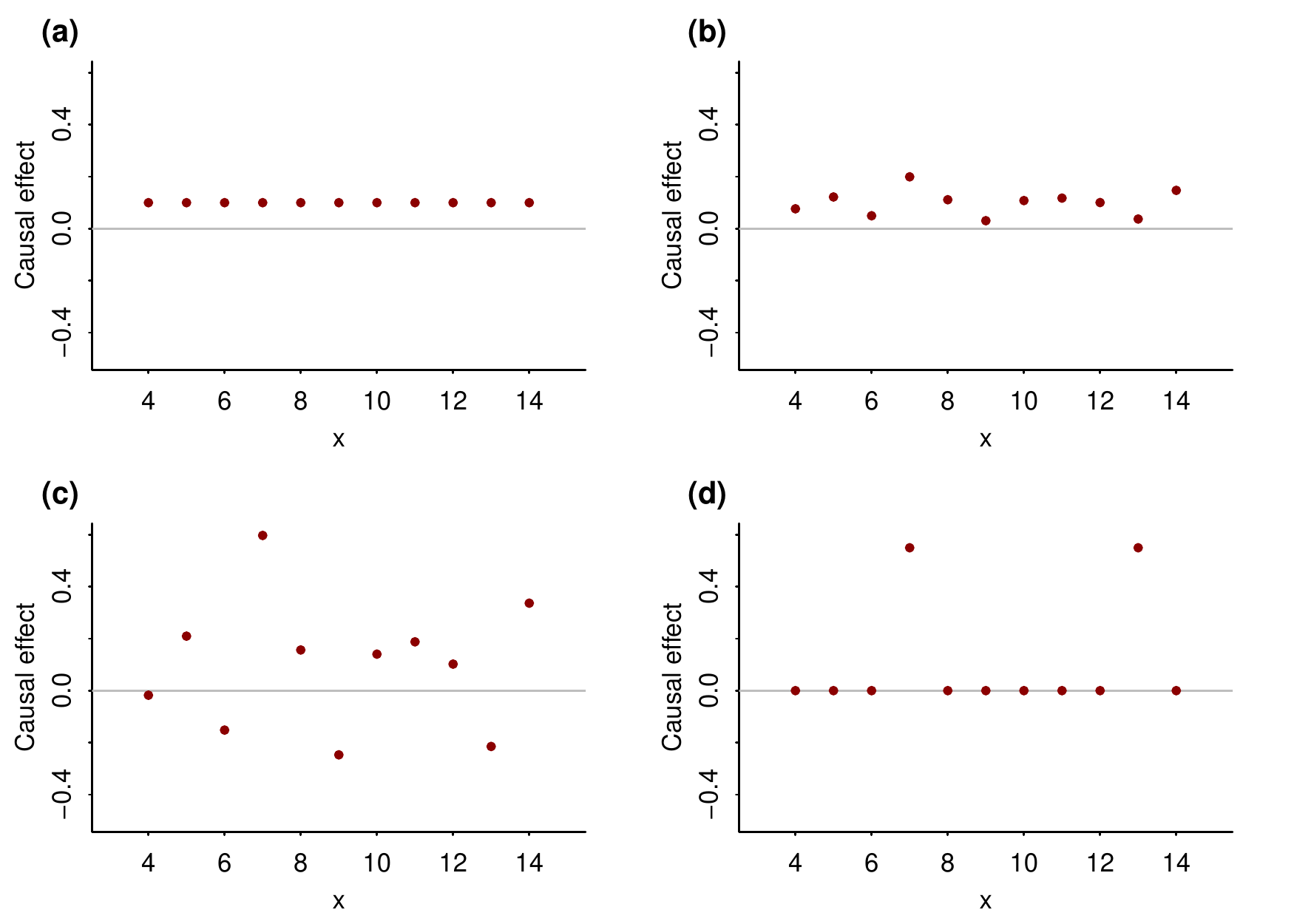}}
\vspace{-.1in}
\caption{\em Four graphs showing different patterns of causal effects, each with average effect of 0.1:  (a) constant effect, (b) low variation, (c) high variation, (d) occasional large effects.}\label{causal_quartet_1}
\end{figure}

Figure \ref{causal_quartet_1} shows four very different scenarios corresponding to an average treatment effect of 0.1.  Figure \ref{causal_quartet_1}a shows the simplest case, often implicitly assumed in discussions of ``the'' treatment effect.  Figure \ref{causal_quartet_1}b shows an effect that is always positive across units but with magnitude varying between 0 and 0.2.  In Figure \ref{causal_quartet_1}c, there is high variation and the effect could be positive or negative at the level of the units. Finally, in Figure \ref{causal_quartet_1}d the treatment effect is usually zero but is high among some small subset of units with nonzero effects.

These plots correspond to four different sorts of real-world situations, and we conjecture that some misunderstanding about effect sizes comes from the habit of thinking about the average effect without considering what that means in the context of variation.  We discuss this in the context of some examples in Section \ref{practical}.

\begin{figure}
\centerline{\includegraphics[width=.8\textwidth]{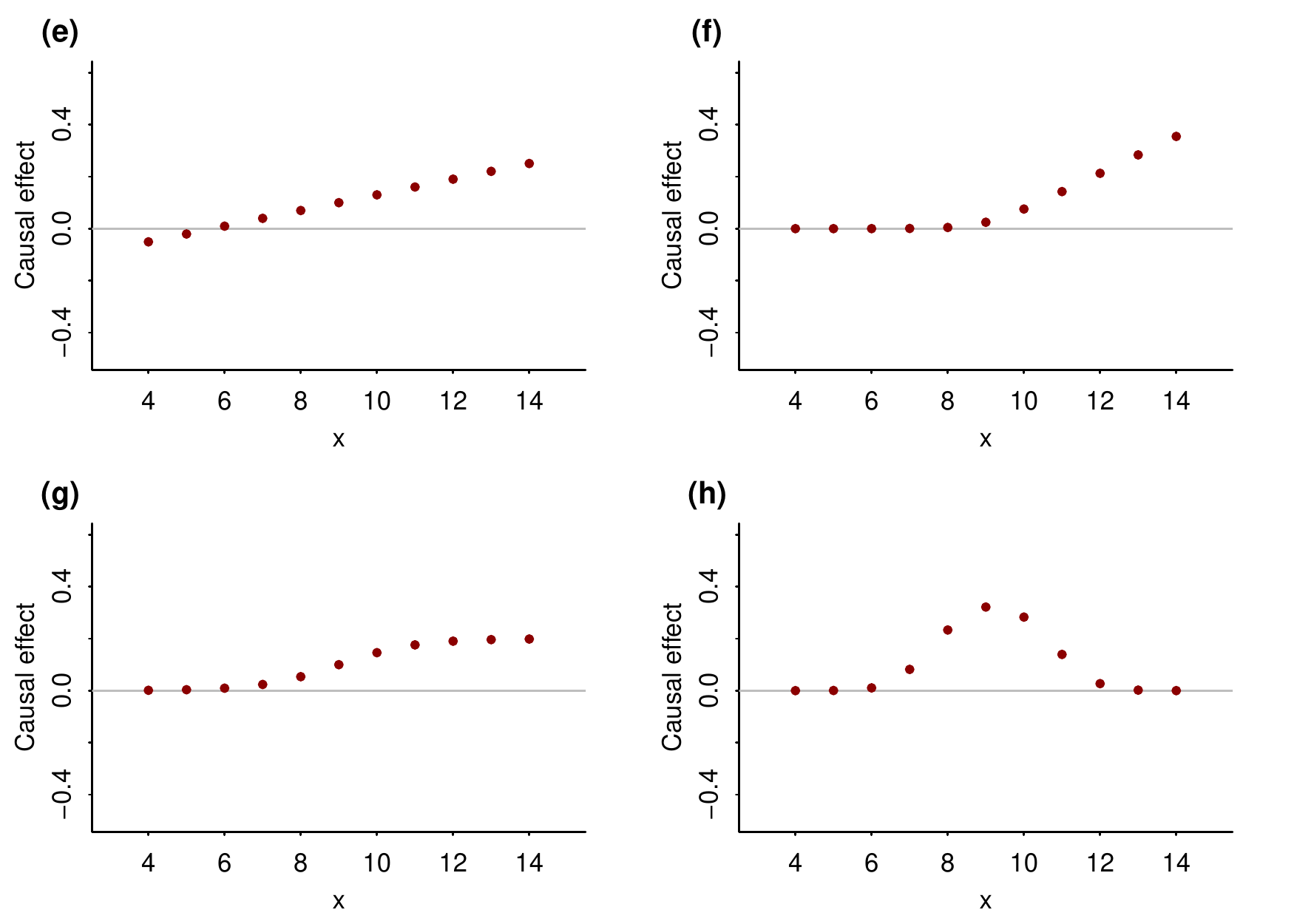}}
\vspace{-.1in}
\caption{\em Four graphs showing patterns of causal effects, each with average effect of 0.1, but varying in different ways as a function of a pre-treatment predictor:  (e) linear interaction, (f) no effect then steady increase, (g) plateau, (h) intermediate zone with large effects.}\label{causal_quartet_2}
\end{figure}

Figure \ref{causal_quartet_2} presents another quartet, this time showing different forms of interaction in which the effect varies systematically as a function of a pre-treatment predictor.  Figure \ref{causal_quartet_2}e shows a linear interaction, which is typically the first model that is fit when researchers go beyond the model of constant effects.  Figure \ref{causal_quartet_2}f illustrates a treatment that is only effective when the pre-treatment variable exceeds some threshold; this could be a training program that requires some minimum level of preparedness of the trainee.  Figure \ref{causal_quartet_2}g adds a plateau, corresponding to the realistic constraint of some maximum effectiveness.  Finally, Figure \ref{causal_quartet_2}h shows a non-monotonic pattern with a ``sweet spot,'' which could arise in a medical treatment that has no effect for the healthiest patients (because they do not need the treatment) or for the sickest (for whom the treatment is too late).

As with Figure \ref{causal_quartet_1}, these are not intended to represent an exhaustive list of possibilities; Figure \ref{causal_quartet_2}e shows the typical assumption made in modeling an interaction, while Figures \ref{causal_quartet_2}f, g, and h represent different sorts of patterns that go beyond what would usually be included in a statistical model.  The first quartet shows different levels and distributions of unpredictable variation; the second represents variation that depends on pre-treatment information.  A realistic setting would include a mix of both.

Our Figures \ref{causal_quartet_1} and \ref{causal_quartet_2} are modeled on the famous correlation quartet of Anscombe (1973):  four scatterplots with the same first and second moments but with much different bivariate patterns.  This quartet is useful for teaching the limitations of the correlation statistic and also stimulating students and researchers to consider alternative models for data.  Later work has explored general approaches to constructing such plots; see Chatterjee and Firat (2007) and Matejka and Fitzmaurice (2017).

\subsection{Plots of observable data}

The big difference between our causal quartets and these earlier correlation quartets is that this earlier work concerned plots of {\em data}, so that departures from the assumed model could be seen directly---hence the title of Anscombe (1973), ``Graphs in statistical analysis''---whereas Figures \ref{causal_quartet_1} and \ref{causal_quartet_2} graph latent {\em causal effects}, which in general cannot directly be observed.  Thus, our plots are conceptual, and their utility to students and researchers is conceptual.  Figures \ref{causal_quartet_1} and \ref{causal_quartet_2} should help in design and analysis of causal studies, both by suggesting ideas for models of treatment effects and as reminders of the limitations of the average causal effect, in the same way that the quartet of Anscombe (1973) dramatized the limitations of the correlation and regression coefficients in descriptive statistics.


\begin{figure}
\centerline{\hspace{.1in}\includegraphics[width=.49\textwidth]{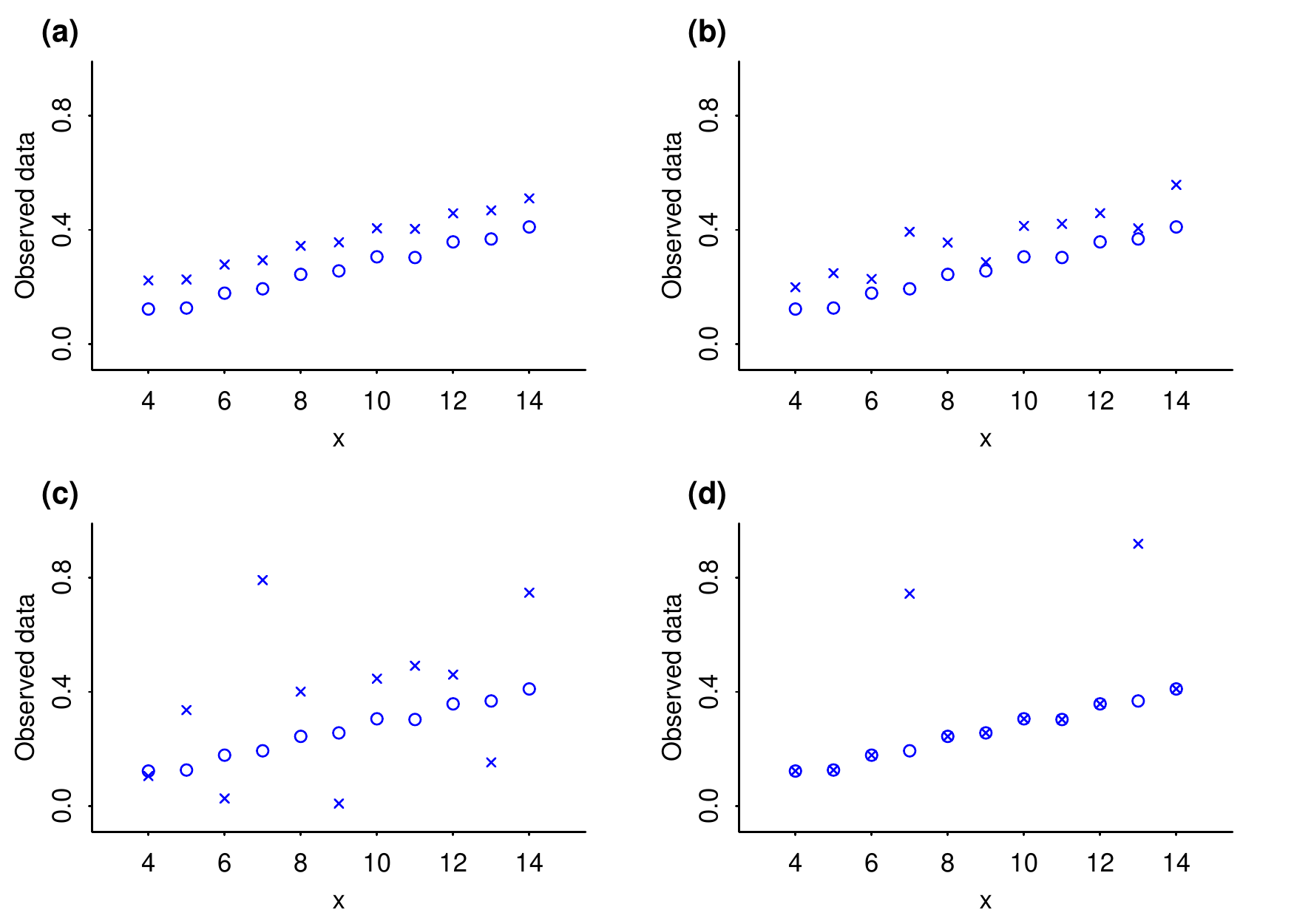}\hspace{.25in}\includegraphics[width=.49\textwidth]{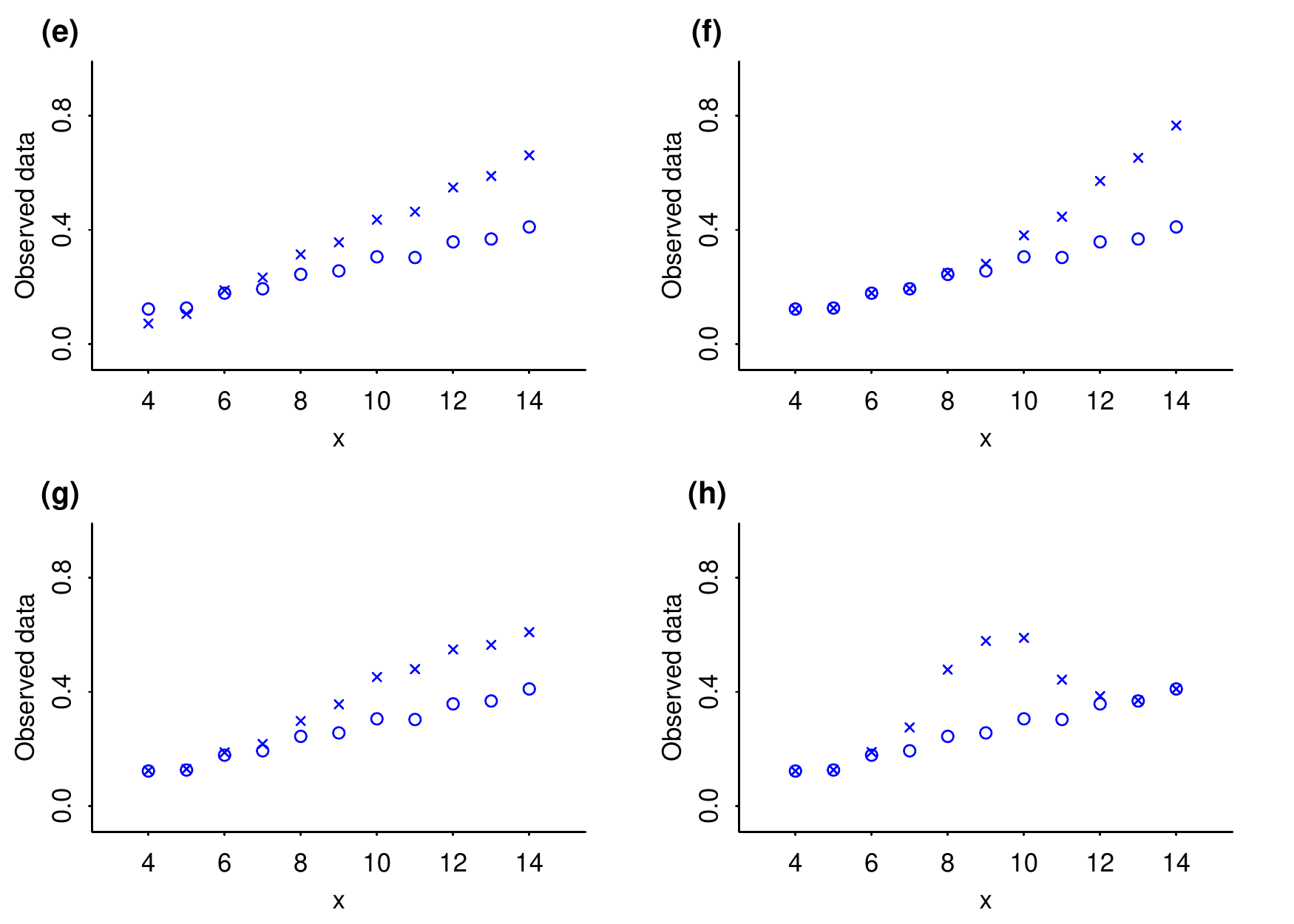}}
\vspace{-.1in}
\caption{\em Two quartets showing different patterns of observable data consistent with the causal effects displayed in Figures \ref{causal_quartet_1} and \ref{causal_quartet_2}.  In each plot, the crosses and circles represent treated and control units, respectively, and the difference between the two is the treatment effect.}\label{causal_quartet_3}
\end{figure}

To better understand these patterns, it can be helpful to visualize them in terms of observable data.  In Figure \ref{causal_quartet_3}, we display graphs of data that are consistent with the effects shown in Figures \ref{causal_quartet_1} and \ref{causal_quartet_2}.  Each of the new plots shows outcomes under treatment and control for the same hypothetical eleven units, with the differences representing the causal effects.

In general, we cannot observe causal effects directly from data: even within-person designs that expose participants to both a control and treatment condition will be affected by factors such as order effects. However, the examples in Figure \ref{causal_quartet_3} give a sense of what data might look like under different patterns of causal effects in the absence of such factors.

\section{Some practical implications of the causal quartets}\label{practical}

We describe some scenarios where causal quartets can help expose problems with assumptions made about the sizes of effects.  This can arise before or after data are collected.

\subsection{Before running a study: Anticipating an effect size}

To design a study one must account for uncertainty in effect sizes. Researchers designing clinical trials often make optimistic assumptions corresponding to high power.  Once we considering variation in the treatment effect, it becomes clear that average effects can be much lower than originally imagined.

We illustrate with an example from Zelner et al.\ (2021) of a doctor designing a trial for an existing drug that he thought could effectively treat high-risk coronavirus patients. He solicited our help to check his sample size calculation that a sample size of $n = 126$ would assure 80\% power under an assumption that the drug increased survival rate by 25 percentage points.  (With 126 people divided evenly split between two groups, the standard error of the difference in proportions is bounded above by $\sqrt{0.5*0.5/63 + 0.5*0.5/63} = 0.089$. To achieve 80\% power requires the value of the effect to be at least 2.8 standard errors from the comparison point of 0, hence, an effect of 0.25 achieves the desired power with $n = 126$.)  When asked how confident he felt about his guess of the effect size, the doctor replied that he thought the effect on these patients would be higher, such that 25 percentage points was a conservative estimate. At the same time, he recognized that the drug might not work.  But when asked what he thought about increasing his sample size so he could detect, for example, a 10 percentage point increase in survival, he replied that this would not be necessary:  he felt confident that if the drug worked, its effect would be large.

It might seem reasonable to suppose that a drug might not be effective but would have a large effect in case of success.  But to stop at this assumption implies a problematic vision of uncertainty.  Suppose, for example, that the survival rate was 30\% among the patients who do not receive this new drug and 55\% among the treatment group.  Here is a hypothetical scenario of what we might expect given 1000 people:
\begin{itemize}
\item 300 would live either way,
\item 450 would die either way,
\item 250 would be saved by the treatment.
\end{itemize}
There are other possibilities consistent with a 25 percentage point average benefit---for example the drug could save 350 people while killing 100---but the point is that once we assume a scenario as we did above, the posited benefit of the drug is not a 25 percentage point benefit for each patient; rather, it's a 100\% benefit for 25\% of the patients.

From that perspective, once we accept the idea that the drug works on some people and not others---or in some comorbidity scenarios and not others---we realize that ``the treatment effect'' in any given study will depend entirely on the patient mix.  There is no underlying number representing the effect of the drug.  Ideally one would like to know what sorts of patients the treatment would help, but in a clinical trial it is enough to show that there is some clear average effect.

Once we consider the treatment effect in the context of variation among patients, as in Figure \ref{causal_quartet_1}c, this can be the first step in a more grounded understanding of effect size.

\subsection{After running a study: Interpreting results}

\subsubsection*{Downgrading an apparently huge effect}

Gertler et al.\ (2013) performed a randomized evaluation of an early-childhood intervention program, yielding an estimate that the program increased adult earnings by 42\%.  This sounds a bit too good be true, even more so when considering it as an {\em average} effect, given that the actual effect must surely vary a lot by person, considering the tortuous path from an intervention at age 4 to earnings at age 24.  A realistic scenario might be some mix of Figure \ref{causal_quartet_1}b and d---effects that are often negligible and can follow a wide range when positive---and Figure \ref{causal_quartet_2}d---an effect that is larger in some intermediate zone.  In any of these cases, we would argue that an average effect of 42\% is hard to believe, given that it would reflect some combination of many effects near zero and some increases in earnings of 100\% or more.

The implication of this reasoning is that the claimed effect is likely to be a wild overestimate---a point that we earlier made on inferential grounds (Gelman, 2018b) but without reference to varying effects.  Combining a realistic sense of the average effect size with an understanding of selection on statistical significance makes it clear that the study had low power and will yield a positively biased estimate (Button et al., 2013).  The framework of nonconstant treatment effects gives us another reason to be skeptical about the claims made for this particular class of interventions.

\subsubsection*{Recognizing that an apparently large effect can be explained as an artifact of noise}

Beall and Tracy (2013) performed two small surveys and found that women were three times as likely to wear red or pink during certain days of their monthly cycle.  This result achieved conventional levels of statistical significance, but this could easily be explained by uncontrolled researcher degrees of freedom; see Simmons, Nelson, and Simonsohn (2011) for a general discussion of this issue and Gelman (2013) in the context of this particular study.

Here, however, we want to focus not on statistical significance but on the reported effect size, which is implausible on its face and even more outlandish when considered as an average effect, once we reflect that the effect will be zero for many people, for example, those who never wear red clothing or those whose clothing choices are restricted because of work.  Even if think it's reasonable to expect a factor-of-3 effect for some women in the study, the average effect including those with no effect would have to be much lower, indeed in this case lower than the uncertainty in the estimated effect 
This implies that the published result, despite its apparent statistical significance, could be explained by a combination of chance and unintentional selection bias.  Indeed, followup studies by these authors and others did not replicate the finding (see, for example, Hone and McCullough, 2020).

Beyond all this, if time of the month does influence clothing choices, we would expect this effect to vary greatly across people and scenarios.  There is no theoretical reason to expect a common direction, hence a pattern such as Figure \ref{causal_quartet_1}c seems likely, to the extent there are effects at all.  Such variation makes it even more difficult to estimate an average treatment effect, as well as implying that any realistic average would be close to zero.

We have used the day-of-cycle and clothing study as an example of the perils of naive interpretation of statistics.  Thinking about varying effects helps us understand why estimating an average effect here is not well motivated:  the problem is not just the lack of successful replication but rather the conceptual framework under which the effect is characterized by a single number or even a single direction.

\subsubsection*{Anticipating the decline effect: Treatments that are less effective in real life}

When designing a medical trial, the first goal is to maximize statistical power.  We say this not cynically but out of a realistic understanding that success---in the form of statistical significance at the conventional level---can be necessary for approval of a drug or procedure, so if you believe your idea is a good one, you want to design your experiment to have a high chance of demonstrating that it works.

Methods of increasing statistical power in an experiment include:  (1) increasing the sample size, (2) improving the accuracy of measurements, (3) including additional pre-treatment predictors, (4) performing within-person comparisons, and (5) increasing the magnitude of the average treatment effect.  Assuming the first four of these steps have been done to the extent possible, one way to achieve the fifth step is to restrict the participants of the study to those for whom the expected effect is as large as possible. 

There is nothing wrong with performing this sort of restriction when designing a study---indeed, it makes a lot of sense in any experiment to focus on scenarios where the signal is highest---and the result should be a higher average treatment effect among participants in the experiment.  When generalizing to a larger population, however, some modeling is necessary conditional on any information used in patient selection.  Thinking about variation in treatment effects makes this clear:  the average effect is not a general parameter; it depends on who is being averaged over.

\section{causalQuartet: An R package for generating causal quartets}\label{package}
To facilitate generating causal quartets, we created an R package called {\tt causalQuartet}.\footnote{\url{https://github.com/jhullman/causalQuartet}}
The package takes as input an average treatment effect and a set of observations \textit{x} (for latent quartets as in Figures \ref{causal_quartet_1} and \ref{causal_quartet_2}) as well as a set of hypothetical control condition values (for observable quartets as in Figure \ref{causal_quartet_3}). The user has the option of specifying additional parameters to control the presentation of the quartets.  

We envision researchers and consumers of research looking at these quartets of hypothetical treatment effects both before and after a study has been run.

For example, in light of well-known problems with overestimated effects when null hypothesis significance testing is applied to low powered studies (e.g., Button et al., 2013), journalists and other consumers of research might benefit from generating quartets to help explore the implications of a claimed effect size. Researchers who wish to interrogate an effect that they have found or that has been reported by their peers, such as when reviewing a paper, might generate a few quartets to support an argument about why an average treatment effect is likely to be an overestimate.

Before running a study, a researcher can use the quartets in deciding what size effect makes sense to target in sample size calculations. The quartets may also be useful in modeling, where they could help promote thinking about how consistent different patterns would be with prior knowledge. For example, a quartet like Figure \ref{causal_quartet_2} could be useful before or after one estimates interactions in a model, to stimulate reflection on the linear assumption. 
By aiming to prompt people to reflect on the sorts of informal expectations they bring to data analysis, the package is similar to prior work by Kim, Reinecke, and Hullman (2017) and Hullman et al.\ (2018), finding that asking users of graphical displays to make predictions about effects before seeing observed data can improve their recall of the data later and their ability to make accurate predictions about new settings. 
For within-person or pre-post treatment designs, a researcher may even want to compare plots of observed subject-specific differences between a treatment and control to a plot like Figure \ref{causal_quartet_3}. However, this should be done with acknowledgment that the observables in Figure \ref{causal_quartet_3} are hypothetical data that are not subject to factors such as order effects that are generally unavoidable in within-person designs.

Another scenario is one we see far too rarely in intervention-oriented empirical research: reflecting on the utility of putting an intervention into practice given an estimated effect. For example, researchers in disciplines ranging from psychology to medicine to economics to computer science often end their interpretation of estimated effects at the average treatment effect. With the help of causal quartets, researchers can instead use the estimate as a jumping off point for discussing the relative utility to be gained from implementing the new intervention under different assumptions about heterogeneity and varying stakes.
\section{Discussion}\label{discussion}

As has been discussed in the judgment and decision making literature, quantities are generally understood comparatively.  Hofman, Goldstein, and Hullman (2020) and Kim, Hofman, and Goldstein (2022) discuss comparisons of effect sizes to inferential or predictive uncertainty.  In the present paper we compared the average causal effect to its variation.

\subsection{Different sources of variation in causal effects}

Figures \ref{causal_quartet_1} and \ref{causal_quartet_2} present this potential variation in an abstract way; in particular applications these can represent variation across experimental units, across situations, and over time, and Figure \ref{causal_quartet_3} can be used to imagine data consistent with such types of variation.  Each type of variation can have applied importance:
\begin{itemize}
\item Variation among people is relevant to policy (for example, personalized medicine) and understanding (for example in psychology, as discussed in Gelman, 2014).
\item Variation across situations is relevant when deciding what ``flavor'' of treatment to do, for example with dosing in pharmacology or treatment levels in traditional agricultural experiments.
\item Variation over time is crucial in settings such as A/B testing where an innovation that has been tested on past data is intended to be applied in the future in an evolving business environment.
\end{itemize}
Variation in effects is itself important, even setting aside inferential and predictive uncertainty in outcomes, that is, even if the true causal effects are known.  That is the point of Figures \ref{causal_quartet_1} and \ref{causal_quartet_2} and the connection to the quartet of Anscombe (1973):  Just as a single number of correlation can represent many sorts of bivariate relationships, so can a single number of average causal effect represent many sorts of causal patterns, even within the simplest setting of a single treatment, a single outcome, and no intermediate variables.

\subsection{Why the causal framework?}

Nothing in this paper so far requires a causal connection.  Instead of talking about heterogeneous treatment effects, we could just as well have referred to variation more generally.  Why, then, are we putting this in a causal framework?  Why ``causal quartets'' rather than ``heterogeneity quartets''?

Most directly, we have seen the problem of unrecognized heterogeneity come up all the time in causal contexts, as in the examples in Section \ref{practical}, and not so much elsewhere.  We think a key reason is that the individual treatment effect is latent.  So it's not possible to make the ``quartet'' plots with raw data. Instead, it's easy for researchers to simply assume the causal effect is constant, or to not think at all about heterogeneity of causal effects, in a way that's harder to do with observable outcomes.  It is the very impossibility of directly drawing the quartets that makes them valuable as conceptual tools.

\subsection{The replication crisis}

The ideas of this paper have has several points of connection to the replication crisis in science:
\begin{itemize}
\item Most immediately, in a world of varying effects, there is no particular interest in testing a null hypothesis of exactly zero effect, and we should be able to move away from the idea that a ``statistical significant'' finding represents something that should replicate; see, e.g., McShane and Gelman (2017).
\item As illustrated in some of the examples in Section \ref{practical}, when we think about how an effect can vary, we often lower our expectations of its average effect, which in turn can make us aware of problems of low power.  For example, if a study is designed under the naive expectation of an effect size of 0.5, but then on reflection we think that an average effect of 0.1 is more plausible, then the study would require 25 times the sample size (or measurements that are 5 times as accurate) in order to maintain the desired power.
\item Moving away from the framing of ``the'' treatment effect helps us think about variation.  Instead of classifying a new study as an exact replication (with the implication that the effect should be the same as in the original study) or a conceptual replication (with the hope that the effect should have the same sign), we can think of the first study and the replication as representing two different collections of participants and situations.
\end{itemize}
As we have argued elsewhere (Gelman, 2015), ``once we accept that treatment effects vary, we move away from the goal of establishing a general scientific truth from a small experiment, and we move toward modeling variation (what Rubin, 1989, calls response surfaces in meta-analysis), situation-dependent traits (as discussed in psychology by Mischel, 1968), and dynamic relations (Emirbayer, 1997). We move away from is-it-there-or-is-it-not-there to a more helpful, contextually informed perspective.''

For example, consider a hypothetical experiment yielding an estimated treatment effect of 0.003 with standard error 0.001, in a setting in which an effect size of 0.1 would be large.  One might first want to dismiss the result as ``statistically significant but not practically significant''---but there are various scenarios under which even a small effect would be notable if its sign is well identified.  In an A/B testing setting in a large company, even an effect of 0.003 could represent many dollars, and in social science we might be interested in the direction of an effect (for example, knowing whether people under stress performed better or worse on a certain task) more than its magnitude.  In such an example, our concern would be that, even if the effect is accurately estimated at 0.003 for this particular experiment, it could easily differ for a new group of people in a different environment.  Perhaps the effect would be $-0.004$ tomorrow, $+0.001$ the next day, and $-0.002$ the day after that.  The relevant comparison is not to the standard error---although that does give us a baseline level of uncertainty---but to changes among people, across scenarios, and over time.  Some of this can be learned from data, other aspects of this variation need to be assumed---but there is generally no good reason to assume that the variation in the treatment effect is zero.

A slightly different argument is that in some applications we really only care about the existence and sign of an effect, not its magnitude:  knowing that an intervention works, even a small amount, could give insight and be relevant for future developments.  But the same problem arises here as before:  there is not necessarily any good reason to believe that a small positive effect in one study will apply elsewhere.  It is not clear how to interpret an average treatment effect, even in a clean randomized experiment, without considering how the effect could vary across people and scenarios and over time.

\subsection{Recommendations for design and analysis}

Looking forward, how should this affect applied research?

To start, with smaller average effect sizes than previously imagined, better designs are needed:  more accurate measurements, better pre-treatment predictors, larger sample size, and within-unit comparisons.

When moving to analysis, interactions are important but hard to estimate with precision.  So when we do include interactions in our model, we should estimate them using regularization and not demand that they attain statistical significance or any other threshold representing near-certainty.

Conversely, when we fit simple models without interactions, we should not expect that the local average treatment effects being estimated to immediately generalize.  Instead, when generalizing we should allow for both predictable and unpredictable variation in effects, even if in doing so we need to hypothesize scales of variation without direct evidence from the data at hand.

When generalizing beyond the observed sample, it is important to account for changes, which can be done by fitting a model accounting for key pre-treatment variables and then poststratifying to estimate the average treatment effect in the new setting (Kennedy and Gelman, 2021).

It is said that in the modern big-data world we should embrace variation and accept uncertainty.  These two steps go together:  modeling of variation is essential for making sense of a world of non-constant treatment effects, but this variation can be difficult to estimate precisely and is sometimes not even identifiable from data, hence the need to accept uncertainty.  Just as the quartet of Anscombe (1973) is a reminder of the limits of correlation that is helpful even when our only readily available analytical tool is linear regression, so we hope the quartets in the present paper can help guide us when thinking about generalizing from local causal identification to future prediction and decision making.

\section*{References}

\noindent

\bibitem Anscombe, F.  J. (1973).  Graphs in statistical analysis. {\em American Statistician} {\bf 27}, 17--21.

\bibitem Baribault, B., Donkin, C., Little, D. R., Trueblood, J. S., Oravecz, Z., Van Ravenzwaaij, D., ... and Vandekerckhove, J. (2018). Metastudies for robust tests of theory. {\em Proceedings of the National Academy of Sciences}, {\bf 115}, 2607--2612.

\bibitem Beall, A. T., and Tracy, J. L. (2013).  Women are more likely to wear red or pink at peak fertility.  {\em Psychological Science} {\bf 24}, 1837--1841.

\bibitem Bryan, C. J., Tipton, E., and Yeager, D. S. (2021).  Behavioural science is unlikely to change the world without a heterogeneity revolution. {\em Nature Human Behavior} {\bf 5}, 980--989.

\bibitem Button, K. S., Ioannidis, J. P., Mokrysz, C., Nosek, B. A., Flint, J., Robinson, E. S., and Munafò, M. R. (2013). Power failure: Why small sample size undermines the reliability of neuroscience. {\em Nature Reviews Neuroscience} {\bf 14}, 365--376.
  
\bibitem Chatterjee, S., and Firat, A. (2007).  Generating data with identical statistics but dissimilar graphics: A follow up to the Anscombe dataset. {\em American Statistician} {\bf 61}, 248--254.

\bibitem Dehejia, R., and Wahba, S. (1999).  Causal effects in non-experimental studies:  Re-evaluating the evaluation of training programs. {\em Journal of the American Statistical Association} {\bf 94}, 1053--1062.

\bibitem Emirbayer, M. (1997). Manifesto for a relational sociology. {\em American Journal of Sociology} {\bf 103}, 281--317.

\bibitem Gelman, A. (2013).  Too good to be true.  {\em Slate}, 24 July.
\url{https://slate.com/technology/2013/07/statistics-and-psychology-multiple-comparisons-give-spurious-results.html}
  
\bibitem Gelman, A. (2014).  When there's a lot of variation, it can be a mistake to make statements about ``typical'' attitudes.  Statistical Modeling, Causal Inference, and Social Science, 8 Oct. \url{https://statmodeling.stat.columbia.edu/2014/10/08/theres-lot-variation-can-mistake-make-statements-typical-attitudes/}

\bibitem Gelman, A. (2015).   The connection between varying treatment effects and the crisis of unreplicable research:  A Bayesian perspective. {\em Journal of Management} {\bf 41}, 632--643. 

\bibitem Gelman, A. (2018a).  You need 16 times the sample size to estimate an interaction than to estimate a main effect.  Statistical Modeling, Causal Inference, and Social Science, 15 Mar.  \url{https://statmodeling.stat.columbia.edu/2018/03/15/need-16-times-sample-size-estimate-interaction-estimate-main-effect/}

\bibitem Gelman, A. (2018b).  The failure of null hypothesis significance testing when studying incremental changes, and what to do about it. {\em Personality and Social Psychology Bulletin} {\bf 44}, 16--23. 

\bibitem Gertler, P., Heckman, J., Pinto, R., Zanolini, A., Vermeerch, C., Walker, S., Chang, S. M., and Grantham-McGregor, S. (2013).  Labor market returns to early childhood stimulation intervention in Jamaica.  Institute for Research on Labor and Employment working paper \#142-13.

\bibitem Hill, J. L. (2011).  Bayesian nonparametric modeling for causal inference.  {\em Journal of Computational and Graphical Statistics} {\bf 20}, 217--240.

\bibitem Hofman, J. M., Goldstein, D. G., and Hullman, J. (2020).  How visualizing inferential uncertainty can mislead readers about treatment effects in scientific results.  {\em Proceedings of the 2020 ACM Conference on Human Factors in Computing Systems (CHI '20)}, 327.

\bibitem Hone, L. S. E., and McCullough, M. E. (2020).  Are women more likely to wear red and pink at peak fertility?  What about on cold days?  Conceptual, close, and extended replications with novel clothing colour measures.  {\em British Journal of Social Psychology} {\bf 59}, 945--964.

\bibitem Hullman, J., Kay, M., Kim, Y., Shrestha, S. (2017). Imagining replications: Graphical prediction \& discrete visualizations improve recall \& estimation of effect uncertainty. {\em IEEE Transactions on Visualization and Computer Graphics} {\bf 24}, 446--456.

\bibitem Imbens, G. W., and Angrist, J. (1994).  Identification and estimation of local average treatment effects.  {\em Econometrica} {\bf 62}, 467--475.

\bibitem Kennedy, L., and Gelman, A. (2021).  Know your population and know your model: Using model-based regression and poststratification to generalize findings beyond the observed sample.  {\em Psychological Methods} {\bf 26}, 547--558.

\bibitem Kim, Y., Hofman, J. M., and Goldstein, D. G. (2022).  Putting scientific results in perspective: Improving the communication of standardized effect sizes.  {\em Proceedings of the 2022 ACM Conference on Human Factors in Computing Systems (CHI '22)}, 625.

\bibitem Kim, Y., Reinecke, K., and Hullman, J. (2017). Explaining the gap: Visualizing one's predictions improves recall and comprehension of data. {\em Proceedings of the 2017 CHI Conference on Human Factors in Computing Systems},  1375--1386.

\bibitem Matejka, J., and Fitzmaurice, G. (2017).  Same stats, different graphs: Generating datasets with varied appearance and identical statistics through simulated annealing. {\em Proceedings of the 2017 CHI Conference on Human Factors in Computing Systems (CHI '17)}, 1290--1294.

\bibitem McShane, B. B., and Gelman, A. (2017). Abandon statistical significance. {\em Nature} {\bf 551}, 558.

\bibitem Mischel, W. (1968).  {\em Personality and Assessment}. New York: Wiley.

\bibitem Rubin, D. B. (1989). A new perspective on meta-analysis. In {\em The Future of Meta-Analysis}, ed.\ K. W. Wachter and M. L. Straff, 155--165.  New York: Russell Sage Foundation.

\bibitem Simmons, J., Nelson, L., and Simonsohn, U. (2011). False-positive psychology: Undisclosed flexibility in data collection and analysis allow presenting anything as significant. {\em Psychological Science} {\bf 22}, 1359--1366.

\bibitem Wager, S., and Athey, S. (2018).  Estimation and inference of heterogeneous treatment effects using random forests. {\em  Journal of the American Statistical Association} {\bf 113}, 1228--1242.

\bibitem Yarkoni, T. (2022). The generalizability crisis. {\em Behavioral and Brain Sciences} {\bf 45}, E1. 

\bibitem Zelner, J., Riou, J., Etzioni, R., and Gelman, A. (2021).  Accounting for uncertainty during a pandemic. {\em Patterns} {\bf 2}, 100310.

\end{document}